\documentclass[twocolumn,twocolappendix]{aastex631}

\usepackage{amssymb}
\usepackage{latexsym}
\usepackage{amsmath}
\usepackage{tikz}
\usepackage{soul}
\usepackage{graphicx}

\shorttitle{A Co-Scaling Grid for Athena++}
\shortauthors{Habegger \& Heitsch}

\graphicspath{{./}{figures/}}

\begin{document}
\bibliographystyle{aasjournal}

\title{A Co-Scaling Grid for Athena++}

\author{Roark Habegger}
\affiliation{University of Wisconsin, Dept. of Astronomy,  Madison 53705, US }

\author{Fabian Heitsch}
\affiliation{University of North Carolina, Dept. of Physics and Astronomy, Chapel Hill, NC 27599, US}

\newcommand{\revise}[1]{%
{#1} %
}

\begin{abstract}
We present a co-scaling grid formalism and its implementation in the magnetohydrodynamics code Athena++. The formalism relies on flow symmetries in astrophysical problems involving expansion, contraction, and center-of-mass motion. The grid is evolved at the same time order as the fluid variables. The user specifies grid evolution laws, which can be independent of the fluid motion. Applying our implementation to standard hydrodynamic test cases leads to improved results and higher efficiency, compared to the fixed-grid solutions.
\end{abstract}


\section{Introduction \label{sec:Intro}}
Many astrophysical phenomena involve evolution in length scale over orders of magnitude. Supernova and kilonova ejecta expand from stellar radii to parsecs before fully mixing with the interstellar gas \citep{Ostriker1988,Montes2016,Metzger2019}. Ionization and stellar-wind driven supershells expand from the scales of a small stellar association to $10$s of parsecs \citep{McCray1987}, impacting the ambient gas and possibly triggering star formation \citep{Elmegreen1977}. Protostellar collapse occurs from parsec-scales to a few astronomical units. 

Numerical modeling of phenomena evolving over a large range of scales requires methods capable of adapting the scale of the spatial discretization. Lagrangian methods such as smoothed particle hydrodynamics \citep{Monaghan1992} achieve this by following fluid elements defined by a fixed mass rather than a fixed volume. Eulerian methods - the choice for many applications because conservation laws are more easily realized - require adaptive mesh algorithms to cover similar scale ranges as smoothed particle hydrodynamics \citep{Krumholz2007,Klein2017}. Lagrangian remapping combines elements of Lagrangian methods with those of Eulerian ones, evolving fluid elements in a Lagrangian frame, but continually re-mapping the motion on to an Eulerian grid \citep{Lufkin1993}. A more recent development are moving-mesh codes, solving flux-conservative problems on meshes that move with the fluid in a Lagrangian fashion \citep{Hopkins2015,2010Springel}, preserving the strength of finite volume methods over smoothed particle hydrodynamics for some applications \citep{Heitsch2006}. 

A conceptually simpler alternative to moving-mesh codes exploits possible symmetries in an astrophysical problem. \revise{The relativistic hydrodynamic codes JET and DISCO have shown the effectiveness of limiting mesh movement to a particular direction in cylindrical coordinates \citep{Duffell2013,Duffell2016}. While symmetries appear during spherical and cylindrical expansion and contraction, they also appear as `co-moving' motion in a single Cartesian direction.} For problems involving a drastic change in spatial scale, a co-scaling grid can be more efficient than adaptive mesh-refinement techniques \citep{Roepke2005}. If uniformity of dissipative properties is relevant, such as for problems involving turbulent transport, a co-expanding grid may be preferable over adaptive mesh refinement. \revise{ To make use of these advantages for co-scaling grids, we implemented the method in the Eulerian grid code Athena++ \citep{Stone2020}.}\footnote{\href{https://github.com/roarkhabegger/athena-TimeDependentGrid}{https://github.com/roarkhabegger/athena-TimeDependentGrid}}

The grid can be co-moving or rescaled, in both cases retaining the initial cell aspect ratio. The grid evolution is integrated at the same time order as the fluid variables. The motion of cell-walls necessitates additional wall flux terms when updating the fluid variables. The time dependence of the grid scaling is defined by a user-specified function. The co-scaling grid can be combined with the adaptive mesh capabilities of Athena++. 

The method improves results of standard test cases. Here, we include the Sod shock tube test and a spherical blast wave test. The sphericity of multi-dimensional blast-waves is preserved on Cartesian grids by a factor of $3$ better than for fixed-grid simulations. While the implementation is a factor \revise{$\sim 1.1$} slower across resolutions and processor number than the stock version of Athena++ for standard hydrodynamics, the advantage of the co-scaling grid lies in its ability to cover spatial (and thus dynamical) ranges over orders of magnitude, resulting in a net efficiency gain.


\section{Formalism \label{sec:Form}}
Eulerian, ideal magnetohydrodynamics solve the  conservation laws
\begin{equation}
   \frac{\partial \textbf{U}^T(\vec{x},t)}{\partial t}
   = -\vec{\nabla}^T\cdot \overline{\textbf{$\Gamma$}} (\vec{x},t) .
   \label{eqn:PDE}
\end{equation}
The row vector $\textbf{U}^T$ contains the conservative variables.
The matrix $\overline{\textbf{$\Gamma$}}$ has columns with the flux of each conservative quantity. These fluxes have rows corresponding to the various coordinate directions ($\hat{x}_1$, $\hat{x}_2$, and $\hat{x}_3$) \citep{Stone2008}. The length of $\textbf{U}^T$ depends on the physics of the problem.  For ideal MHD, $\textbf{U}^T$ has $8$ components and the matrix $\overline{\textbf{$\Gamma$}}$ has $3$ rows and $8$ columns.  Altogether, the right hand side is the flux divergence. For a Cartesian grid, the matrix $\overline{\textbf{$\Gamma$}}$ has the form
\begin{equation}
  \overline{\textbf{$\Gamma$}} = \textbf{F}^T\hat{x} + \textbf{G}^T\hat{y} + \textbf{H}^T\hat{z}
  \label{eqn:FluxMatrix}
\end{equation}
where each boldfaced vector of conservative variables is the flux of those quantities in the given direction.

\revise{
By integrating Eqn.\,\ref{eqn:PDE} over a discrete volume $\Delta V$, the differential equation becomes an integro-differential equation. For static grids, this equation can be rewritten as an ordinary differential equation for the conservative variables $\textbf{U}$ of each cell, indexed by $(i,j,k)$:
}
\revise{
\begin{multline}
    \frac{d}{dt} \textbf{U}_{i,j,k} = 
        - \frac{1}{\Delta x_{i}}\left( 
        \textbf{F}_{i+\frac{1}{2},j,k} 
        -  \textbf{F}_{i-\frac{1}{2},j,k}\right) \\
        - \frac{1}{\Delta y_{j}}\left( 
        \textbf{G}_{i,j+\frac{1}{2},k}
        -  \textbf{G}_{i,j-\frac{1}{2},k} \right) \\
        - \frac{1}{\Delta z_{k}}\left( 
        \textbf{H}_{i,j,k+\frac{1}{2}}
        -  \textbf{H}_{i,j,k-\frac{1}{2}}\right)
    \label{eqn:StaticUpdate}
\end{multline}
}

\revise{where the conservative variables $\textbf{U}$ are averaged over the cell volume and the flux vectors $\textbf{F}$, $\textbf{G}$, $\textbf{H}$ are averaged over a cell wall (see \cite{Stone2020,Felker2018}).} In Eqn.~\ref{eqn:StaticUpdate}, we have removed the row and column vector notation since there are no vector operations left. Therefore, either case (row or column vectors) would satisfy the equation as it is shown. We show a more detailed derivation of Eqn.\,\ref{eqn:StaticUpdate} in Appendix A.

A critical part of the discrete averaging leading to Eqn.~\ref{eqn:StaticUpdate} is moving the time derivative out of the volume integral on the left hand side of Eqn.\,\ref{eqn:PDE} (see Appendix A). The justification for that step is the Reynolds Transport Theorem for a quantity $f$ over a volume $V$ and boundary $B$,
\begin{equation}
    \frac{d}{dt}\int_V dV \, f
    = \int_V dV \, \frac{\partial f}{\partial t} 
    + \int_{B} dA \, (\vec{w} \cdot \hat{n}) f .
    \label{eqn:RTT}
\end{equation}
For a static grid, the velocity $\vec{w}$ of the boundary is $0$, so Eqn.~\ref{eqn:RTT} reduces to
\begin{equation}
    \frac{d}{dt}\int_V dV \,  f 
    = \int_V dV \, \frac{\partial f}{\partial t} ,
    \label{eqn:RTTStatic}
\end{equation}
allowing the time derivative to be moved in and out of any volume integral.

A co-scaling grid will lead to additional fluxes due to moving cell walls, rendering the surface integral in Eqn. \ref{eqn:RTT} non-zero \citep{springel2011}.  \revise{The differential equation now reads (see App.~A for details)
\begin{multline}
    \frac{d}{dt} \textbf{U}_{i,j,k} = \\
        - \frac{1}{\Delta x_{i}}\left( 
        \textbf{F}_{i+\frac{1}{2},j,k} - \textbf{F}_{i-\frac{1}{2},j,k} 
        - \textbf{W}_{i+\frac{1}{2},j,k} + \textbf{W}_{i-\frac{1}{2},j,k}\right) \\
        - \frac{1}{\Delta y_{j}}\left( 
        \textbf{G}_{i,j+\frac{1}{2},k} -  \textbf{G}_{i,j-\frac{1}{2},k} 
        - \textbf{V}_{i,j+\frac{1}{2},k} + \textbf{V}_{i,j-\frac{1}{2},k} \right) \\
        - \frac{1}{\Delta z_{k}}\left( 
        \textbf{H}_{i,j,k+\frac{1}{2}} - \textbf{H}_{i,j,k-\frac{1}{2}}
        - \textbf{S}_{i,j,k+\frac{1}{2}} + \textbf{S}_{i,j,k-\frac{1}{2}} \right)
    \label{eqn:DynamicUpdate}
\end{multline}
}
where $\textbf{W}$, $\textbf{V}$, $\textbf{S}$ are the volume-averaged wall fluxes in the various Cartesian coordinate directions.  For example, the average wall flux in the $\hat{x}$~direction is given by the integral 
\revise{
\begin{multline}
    \textbf{W}_{i+\frac{1}{2},j,k} =
    \frac{1}{\Delta y_j \Delta z_k} \\
    \iint dy dz \,
    \left[ w_x(x_{i+\frac{1}{2}},t) \textbf{U}(x_{i+\frac{1}{2}},y,z,t) \right].
    \label{eqn:WallFlux}
\end{multline}
}

\revise{While implementing the above correction is an important step, there is another correction hidden in Eqn.~\ref{eqn:DynamicUpdate}. The code will use a time integrator to solve the differential equation. Regardless of the particular integrator, the time integration reads}
\revise{
\begin{multline}
    \textbf{U}_{i,j,k}(t^{n+1}) = \textbf{U}_{i,j,k}(t^{n}) \\ 
    + \frac{1}{t^{n+1} - t^n}
    \int_{t^n}^{t^{n+1}} dt \left[ \frac{d}{dt} \textbf{U}_{i,j,k} \right].
    \label{eqn:TimeInt}
\end{multline}}
\revise{
This assumes a static grid. The assumption is hidden in the notation: for a static grid, the volume averages of $\textbf{U}$ are taken over the same volume. To correct for this in the time dependent grid case, we need to change Eqn.~\ref{eqn:TimeInt} to }
\revise{
\begin{multline}
    \textbf{U}_{i,j,k}(t^{n+1}) = \frac{V_{i,j,k}(t^n)}{V_{i,j,k}(t^{n+1})}
    \Bigg[ \textbf{U}_{i,j,k}(t^{n}) \\ 
    + \frac{1}{t^{n+1} - t^n}
    \int_{t^n}^{t^{n+1}} dt \left[ \frac{d}{dt} \textbf{U}_{i,j,k} \right] \Bigg].
    \label{eqn:TimeInt2}
\end{multline}}
\revise{
Here, $V_{i,j,k}(t)$ is the volume of the $(i,j,k)$ cell at time $t$.}

\revise{Thus, a co-moving, co-scaling, or generically time-dependent grid requires two corrections.
The first is to include cell wall movement by using the true flux (Eqn.~\ref{eqn:DynamicUpdate}). The second is to include the change in cell volume, scaling the conserved quantities  (Eqn.~\ref{eqn:TimeInt2}).
}

\noindent

\section{Implementation \label{sec:Impl}}
Athena++ solves Eqn.\,\ref{eqn:PDE} over a static grid \citep{Stone2008,Stone2020}. A co-scaling grid requires the integration of the grid's motion over time, in addition to the integration of the physical variables. After this grid integration, we add corrections to the physical variables in the form of wall fluxes and volume scaling \revise{(Sec.~\ref{ss:wallfluxes}; derived in Sec.~\ref{sec:Form})}. Finally, all coordinate variables need to be updated throughout the full mesh hierarchy, including derived quantities such as cell volumes and areas, and reconstruction coefficients. This requires changes to the task list implemented in Athena++ (Sec.~\ref{ss:tasklist}). 

\subsection{Wall Flux and Volume Change Corrections}\label{ss:wallfluxes}

\begin{figure}
    \centering
    \includegraphics[width=0.48\textwidth]{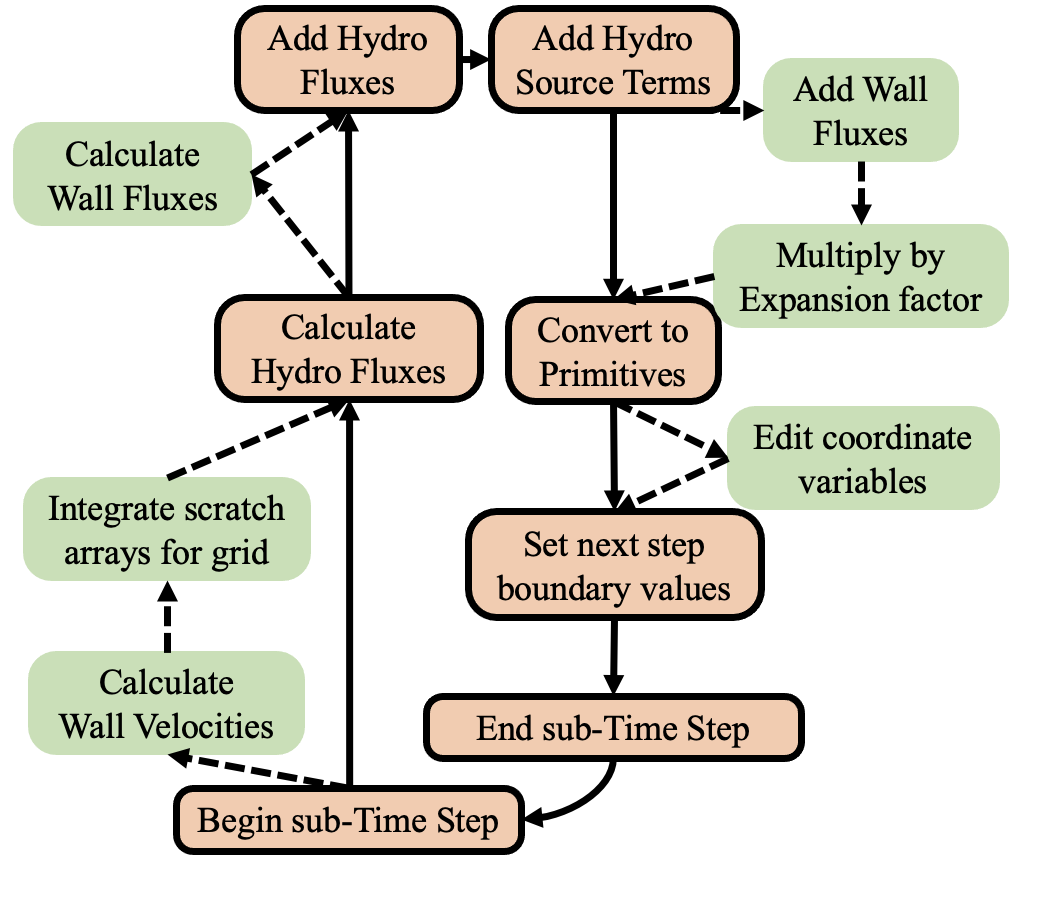}
    \caption{Changes to the task list in Athena++ for the co-scaling grid module. The orange shaded boxes (with black borders) show the normal progression through the task list during a given sub step of a time integrator. The green boxes \revise{(with no borders)} and dashed arrows are the detours necessary for a co-scaling grid. For simplicity, we only show base Athena++ tasks affected by the co-scaling grid. For more detailed flow charts of Athena++ task lists, see \cite{Stone2008,Stone2020}.}
    \label{fig:TaskList}
\end{figure}

To include the corrections to the update equation (Eqn.~\ref{eqn:DynamicUpdate}), we need to approximate the wall flux integral (Eqn.~\ref{eqn:WallFlux}). \revise{Assuming the cell wall's velocity and the conservative quantity are constant on the cell wall,  Eqn.~\ref{eqn:WallFlux} reads 
\begin{equation}
    \textbf{W}_{i+\frac{1}{2},j,k}
    = w_x(x_{i+\frac{1}{2}},t) \textbf{U}(x_{i+\frac{1}{2}},t),
    \label{eqn:1DWallFlux}
\end{equation}
providing a simple definition for the wall flux. }
\revise{
To add these fluxes, we introduce an `expansion' source function. Here, we add the wall fluxes to the conservative variables in the same manner used in the base Athena++ code to add the hydrodynamic fluxes (except we need to consider the difference in sign, see Eqn.~\ref{eqn:DynamicUpdate}). Then, we multiply all conservative variables by the volume expansion factor $V_{i,j,k}(t^n) \left(V_{i,j,k}(t^{n+1}) \right)^{-1}$ (see Eqn.~\ref{eqn:TimeInt2})
With this last step, the conservative variables are fully updated. These steps are shown as a part of Fig.~\ref{fig:TaskList}. }
\newline

\subsection{Task List Changes}\label{ss:tasklist}
Athena++ uses a task list to control and optimize the sequence of operations necessary to solve Eqn.~\ref{eqn:PDE} \citep{Stone2008,Stone2020}. For any time integrator, the code completes the task list for every time \textit{sub-step}. For example, when Athena++ runs with a 4th order time integrator, the task list completes 4 times during a time step, once for each sub-step. Each loop through the list is slightly different, because Athena++ uses minimum-register time integrator methods \citep{Ketcheson2010}. By incorporating the co-scaling grid integration into this task list, the implementation works for any time integrator available in Athena++.

The co-scaling grid requires additional tasks during a sub-step.  The first is an evaluation of the user-prescribed velocity function for every cell wall of the grid. The second task takes those stored velocities and integrates the grid, over time, to determine where each cell wall will be at the end of the sub-step. The third edits the stored coordinates to reflect the change in location of each cell wall. The other detour boxes in Fig.\,\ref{fig:TaskList} regarding wall flux calculation and correcting the conservative variables are implemented within other tasks in the task list. 

The first two new tasks are executed before any hydrodynamic or magnetic field calculations, since the wall velocity function only uses information from the previous time sub-step (see Fig.\,\ref{fig:TaskList}).

We update the coordinate grid after the conservative and primitive variables have been fully updated, and before boundary values are calculated for the next time sub-step. As a result, all variables (grid and physical) are fully updated and available for output or the next integration step. 


\begin{figure}
    \centering
    \includegraphics[width=0.48\textwidth]{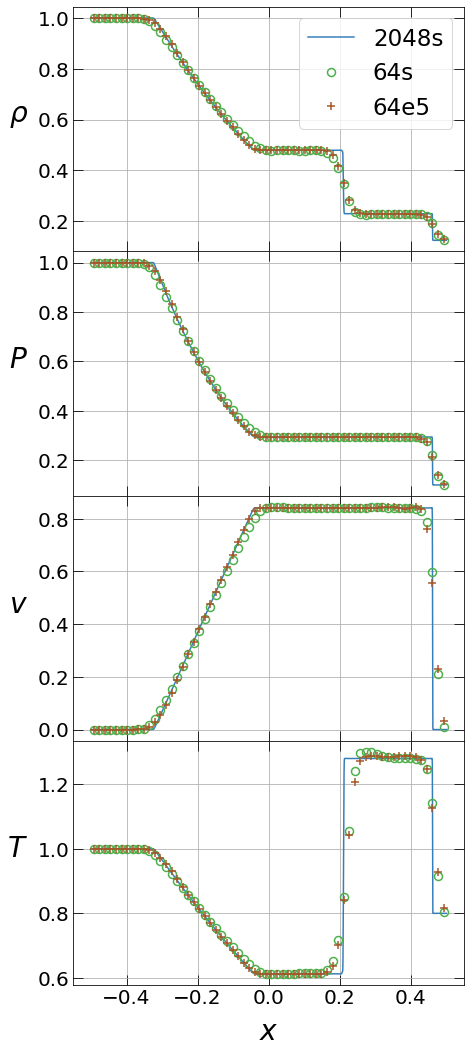}
    \caption{The primitive variables at for the Sod shock tube at $t=0.25$. The line corresponds to a resolution of $N=2048$ on a static grid, and the green circles represent a static grid model at $N=64$ cells. The corresponding co-scaling grid model at $N=64$ is indicated by the brown plus markers. The latter starts out with a domain $[-0.1,0.1]$ and expands to the domain shown, by a factor of $5$.}
    \label{fig:Sod}
\end{figure}

\section{Tests \label{sec:Test}}
We assess accuracy and stability of the co-scaling grid implementation with the 1D Sod shock tube \citep{1978Sod,Stone2008}, and with the 2D cylindrical blast wave. For the latter, we compare the accuracy and computational cost of our implementation to the equivalent static grid simulation. To highlight the applicability of the co-scaling grid to astrophysical problems, we finish with the evolution of a blast waves from free expansion to the Sedov-Taylor phase.

Each test uses a gas with a specific heat capacity ratio $\gamma=\frac{5}{3}$.

\subsection{Sod Shock Test}

The Sod shock tube is a popular test of a numerical code's accuracy and stability  \citep{1978Sod,Stone2008} by comparing the numerical solution to the corresponding Riemann problem (e.g. \cite{Toro2019}). The initial conditions of the test are a density and pressure discontinuity at the origin with $0$ velocity throughout the simulation. The test uses Cartesian coordinates in one dimension.

The left side of the initial discontinuity has density $\rho_l = 1$ and pressure $P_l = 1.0$, whereas the right side has density $\rho_r = 0.125$ and pressure $P_r = 0.1$ \citep{1978Sod}.

Fig.\ref{fig:Sod} compares a co-scaling grid simulation (64e5) with a static grid at the same number of grid points (64s). We also show a higher resolution static grid simulation (2048s) as an approximation of the analytic solution.

\begin{figure}
    \centering
    \caption{Time evolution of the Sod shock tube. The top row shows the density profile for the three models 64e5, 64s, and 2048s, the bottom row the normalized residuals with respect to the 2048s model. Large errors at the discontinuities arise because the discontinuities cannot be resolved physically. For smooth flow regions, the co-scaling grid solution approximates the $2048s$ simulation more closely.}
    \includegraphics[width=\columnwidth]{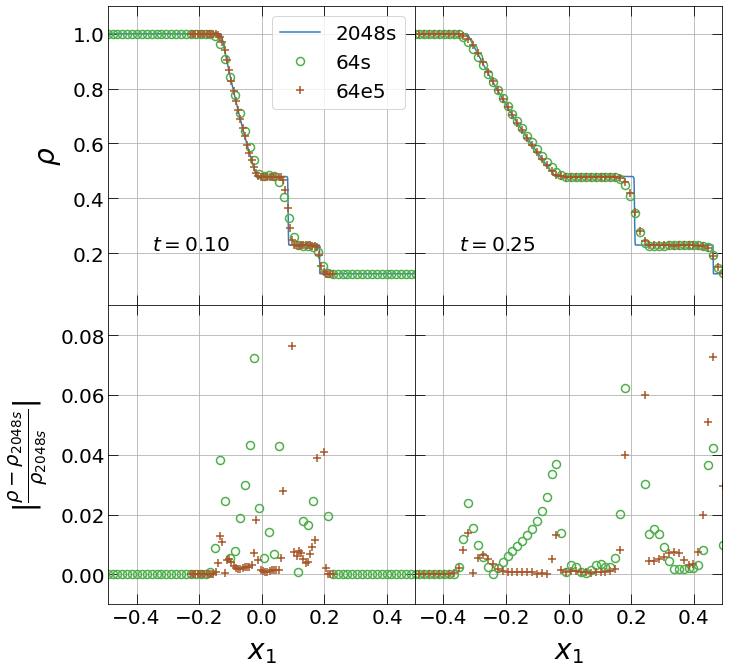}
    \label{fig:SodTime}
\end{figure}

For the co-scaling grid, the domain initially extends over $-0.1\leq x \leq 0.1$ and expands by a factor of $5$, reaching the static grid domain size of $-0.5\leq x \leq 0.5$ at $t=0.25$. Thus, the final output of the simulations can be directly compared (see Fig.\,\ref{fig:SodTime}). The expanding grid keeps the cell size uniform.

In terms of code validation, the solution for the co-scaling grid is consistent with the analytic solution (Fig.\,\ref{fig:Sod}). Specifically for $x\leq 0$, the co-scaling grid approximates the analytic solution more closely than the static grid. Large differences are expected at the discontinuities (see also Fig.\,\ref{fig:SodTime}), since the discontinuities cannot be resolved -- slopes just get steeper with increasing resolution. Generally, slopes are slightly steeper for the co-scaling grid, suggesting that tracking the three waves with the co-scaling grid improves the accuracy of the solution. \revise{Fig.\,\ref{fig:SodTime} indicates the co-scaling grid is more accurate over the evolution of the test, with steeper discontinuities and more accurate shock location. }

\revise{The expanding and static simulations used a Piecewise-Linear Reconstruction (PLM) method. The Piecewise-Parabolic Reconstruction (PPM) method is known to cause oscillations in the velocity \citep{Lee2011}. The PPM method combined with the co-scaling grid results in higher oscillations than when using the static grid. Since the co-scaling grid requires more time steps and thus more reconstructions, oscillations can reach higher amplitudes. 
}


\subsection{1D Blast Wave Test}
Our second test is the one-dimensional blast wave in spherical coordinates. This allows us to \revise{check the volume expansion correction in other coordinate systems.} \revise{The initial condition consists of an inner region with $r<1.0$, which is over-pressured by a factor of $ 10^4$ and over-dense by a factor of $10^3$ with respect to the ambient medium ($\rho_\mathrm{amb} = 0.1$, $P_\mathrm{amb} = 1.0$).}

\begin{figure}
    \includegraphics[width=\columnwidth]{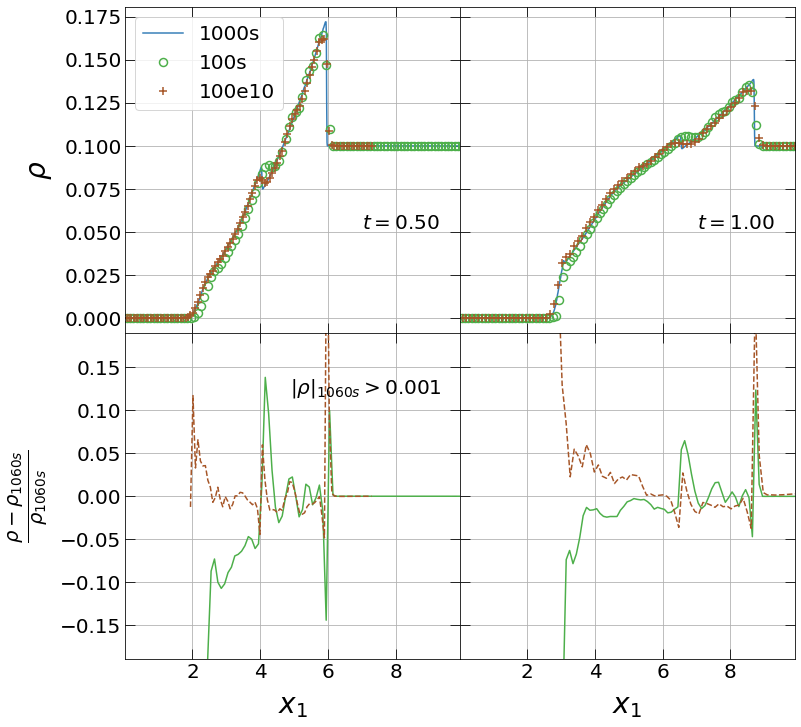}
    \caption{\revise{Density and normalized residual profiles at two times for the spherical 1D blast wave at a resolution of $100$ cells. Residuals are calculated with respect to the static high resolution simulation at $1000$ grid points (blue solid line). Discontinuities introduce large residuals. The expanding grid initially matches the resolution of high resolution profile, leading to more accurate shock locations. }}
    \label{fig:BlastDens}
\end{figure}

\revise{As for the Sod shock tube, we compare three models, one at high resolution ($1000$ grid points), and the fixed and expanding grid models at $100$ grid points each.} Fig.\,\ref{fig:BlastDens} summarizes the results, showing the density profile and normalized residuals for two time instances. While discontinuities introduce large residuals, the shock position is more accurately traced by the co-scaling grid model -- for the fixed-grid model, the shock position leads compared to the high-resolution model. 

\revise{Early in the expanding simulation, the peaks are significantly more resolved than in the static grid simulation. The sharpness of discontinuities plays an important role in radiative losses. Therefore, simulations with radiative losses will be more accurate if they use a co-scaling grid.}

\subsection{2D Blast Wave Test}\label{ss:2dblast}
We test the multi-dimensional performance of the co-scaling grid via the 2D blast wave, both for cylindrical and Cartesian coordinates. Cartesian coordinates introduce directionally dependent numerical diffusion, since the resolution is effectively lower along the diagonals by a factor of $\sqrt{2}$.

\begin{figure}[h]
  \includegraphics[width=\columnwidth]{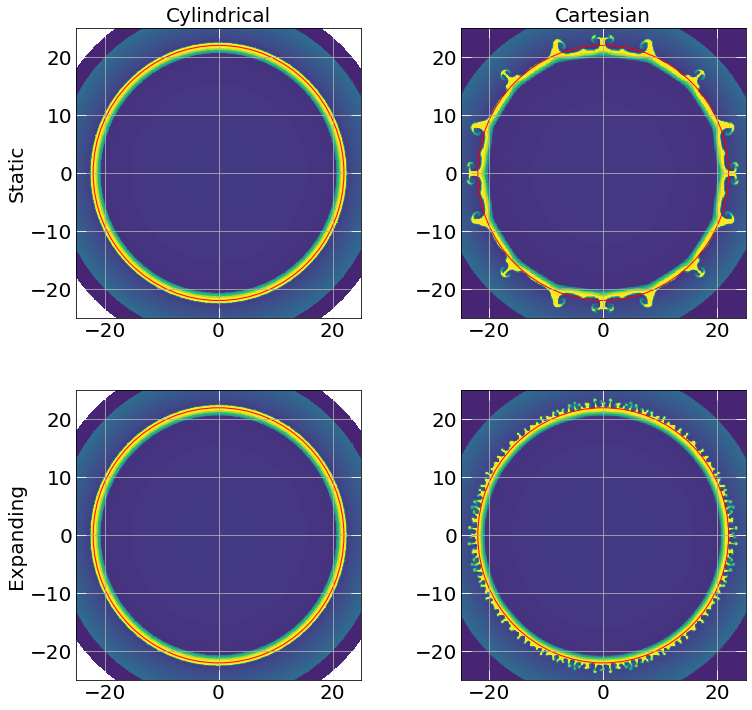}
  \caption{Density maps of four two-dimensional cylindrical blast wave models. For reference, a red circle is plotted at $r=22.5$ in all the maps. This is the radial coordinate of the peak in density for the cylindrical simulations. Models on cylindrical grids (left, $n_r=512$) are nearly indistinguishable. The strength of the co-scaling grid (bottom row) becomes clear when comparing Cartesian grid models (right column, $n_x=n_y=512$). Rayleigh-Taylor fingers triggered by the discretization are nearly uniformly distributed for the co-scaling grid.
  \label{fig:2DBlast}}
\end{figure}

Fig.\,\ref{fig:2DBlast} compares four 2D blast wave models. Cylindrical models on a grid with resolution $(n_r,n_\theta)=(512,64)$ are shown on the left, Cartesian ones at linear resolution of $512$ on the right, while the top row shows static grid models, and the bottom row co-scaling grid ones. We track the shell to determine the grid expansion rate required to keep the blast wave within the simulation domain. 

\revise{The cylindrical models are essentially indistinguishable, as expected. Using cylindrical (instead of spherical) coordinates allows us to more easily compare to the 2D Cartesian case.}

Both Cartesian cases suffer from directionally dependent numerical diffusion, yet, the co-scaling grid achieves a more spherical solution. \revise{We expect Rayleigh-Taylor fingers to be triggered at the grid scale. Since the expanding grid starts with a smaller scale, the instabilities are seeded at a smaller scale, leading to a more uniform distribution of Rayleigh-Taylor fingers and a more circular appearance of the blast wave.}  

\begin{figure}[h]
    \centering
    \includegraphics[width=0.48\textwidth]{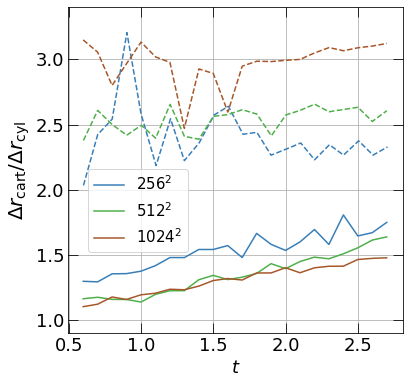}
    \caption{Deviation from circular shape (eqn.~\ref{eqn:estr}), against time, for the blast wave on a Cartesian grid. 
    The dashed lines are static grid simulations and the solid lines are expanding grid simulations. At the same resolution the co-scaling grid simulations produce shells which are more uniformly circular.}
    \label{fig:Acc}
\end{figure}

Fig.\,\ref{fig:Acc} provides a more detailed view of the deviation from sphericity. The vertical axis measures the difference between the largest outer and smallest inner radius of the shell $\Delta r$ for the Cartesian grid, normalized by the same quantity for the cylindrical grid. An effective shell thickness ratio of
\begin{equation}
{\cal{R}} \equiv \frac{\Delta r_{cart}}{\Delta r_{cyl}}=1\label{eqn:estr}
\end{equation}
indicates a perfectly circular ring. The less circular the shell, the larger $\Delta r_{cart}$ will become, and thus $\mathcal{R} > 1$. Results for the co-scaling grid (solid lines) improve with higher resolution.  With time, deviations from circularity grow because the discretization leads to Rayleigh-Taylor instabilities (see Fig.\,\ref{fig:2DBlast}). The effective shell thickness ratios ${\cal{R}}$ for the static grid (dashed lines) are at least a factor of $2$ larger than for the co-scaling grid and vary substantially with time right from the start of the shell expansion.

\begin{figure}[h]
  \centering
  \includegraphics[width=\columnwidth]{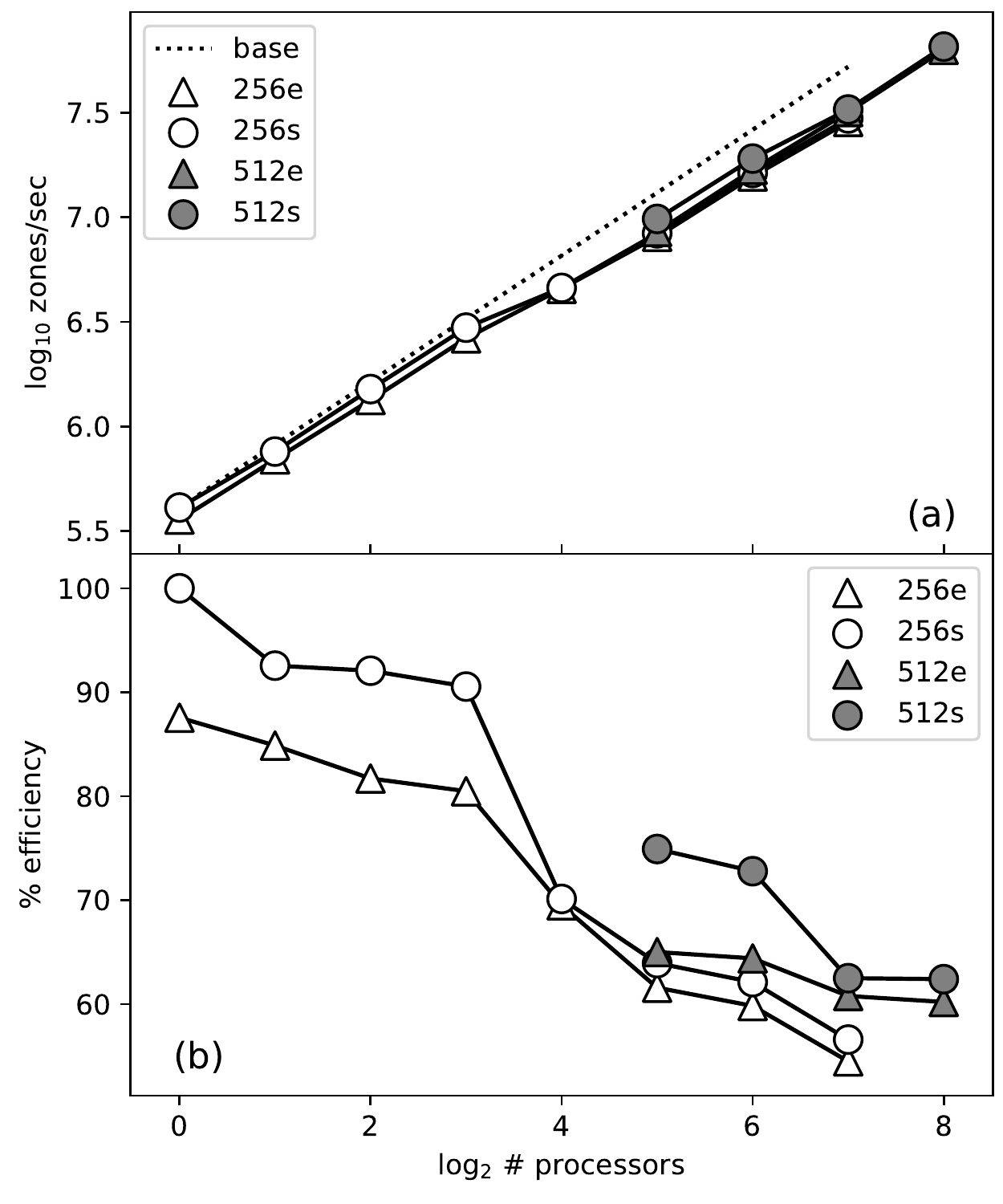}
  \caption{\revise{(a) Speed of the fixed grid (circles) and co-scaling grid (triangles) against processor number, as a strong scaling measure. (b) Efficiency of the fixed and co-scaling grids measured with respect to base efficiency of the fixed grid using $1$ processor. The co-scaling grid tracks the stock version of Athena++, running at $\sim 90$\% of the base speed for small processor numbers, and without perceptible loss for large numbers.}}
  \label{fig:Res}
\end{figure}

\subsection{Performance}

To compare the performance of the stock version of Athena++ with the co-scaling grid implementation (Fig.\,\ref{fig:Res}), we ran the blast wave test in three dimensions in Cartesian coordinates at a resolution of $256^3$ and $512^3$. We use basic hydrodynamics, i.e. no magnetic fields or other additional physics, except for the co-scaling grid. The co-scaling grid implementation tracks the stock version closely, running at $90$\% of the base speed for small processor numbers, and without perceptible loss for large processor numbers. This behavior extends to the $512^3$ resolution, and thus seems resolution -independent, demonstrating that our modification to Athena++ is ``minimally invasive". \revise{While the additional steps clearly slow the code down, the speed decrease is offset with increased accuracy of the co-scaling implementation (Sec.~\ref{ss:2dblast})}.

\subsection{Long-Term Blast Wave Evolution}
\begin{figure}[h]
    \centering
    \includegraphics[width=\columnwidth]{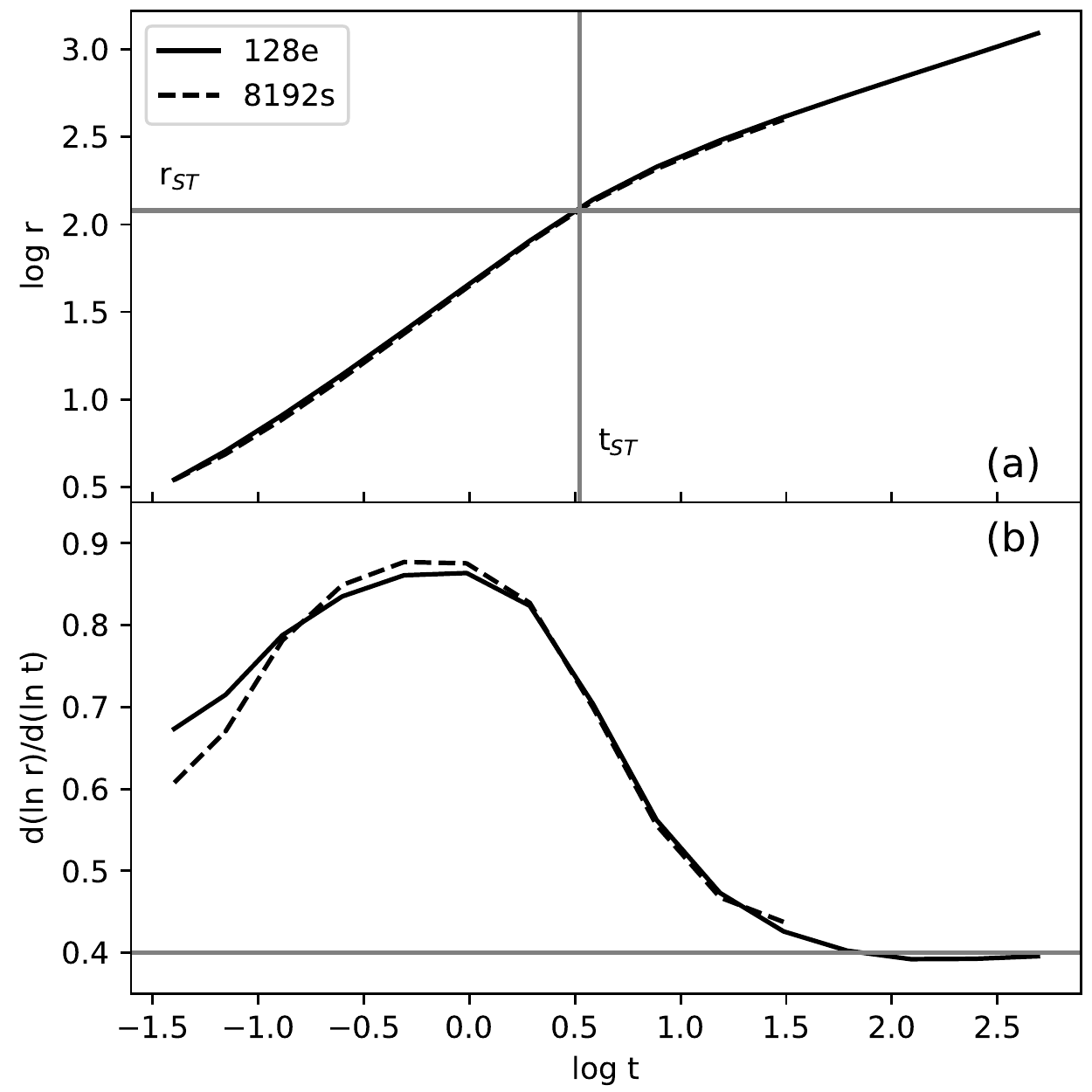}
    \caption{(a) Blast wave radius against time for the fixed grid (8192f) and the expanding grid (128e) model. Grey solid lines indicate the transition radius (eq.~\ref{eq:transrad}) and time. (b) Logarithmic slope of $r(t)$, indicating the transition from the free expansion to the Sedov-Taylor phase ($d(\ln r)/d(\ln t)=0.4$).}
    \label{fig:fest}
\end{figure}

As a final demonstration of the code's capabilities, we follow the evolution of a point explosion from free-expansion to the Sedov-Taylor phase as in a supernova or kilonova remnant \citep{Ostriker1988,Montes2016}. During free expansion, the velocity is constant and the radius scales as $r_s\propto t$. When the ejecta mass reaches the mass of the swept-up ambient gas, the blast wave enters the energy-conserving Sedov-Taylor phase with $r_s\propto t^{2/5}$. Once radiative losses become dominant, the snow-plow (momentum-conserving) phase is reached. \revise{ Here, we only consider the first two phases, leaving the implementation of radiative losses for a later contribution.} The explosion is initialized with a total energy of $E=10^8$, with kinetic energy $E_{kin}=0.99E$. The ejecta density is set to $\rho_e=10^4$ within a radius of $r_s(0)=1.2$. The ambient density and pressure are $\rho_a=10^{-2}$ and $P_a=10^{-7}$. These values result in a transition radius between free expansion and Sedov-Taylor phase of 
\begin{equation}
  r_{ST}=r_s(0)\left(\frac{\rho_e}{\rho_a}\right)^{1/3}=120. \label{eq:transrad}
\end{equation}
We implemented the test for the expanding grid at $128$ grid points and for the fixed grid at an approximate equivalent of $8192$ points (Fig.\,\ref{fig:fest}). Results agree with the analytical estimate for both implementations. The advantage of the expanding grid is obvious -- it can follow the evolution to arbitrary time values. A more sophisticated implementation would include radiative losses to allow the blast wave to enter the snow-plow phase. 


\section{Discussion}

The co-scaling grid implementation provides a generalization of the co-expanding grid formalism of \citep{Roepke2005}, applicable to expanding, contracting, or comoving frames. Its closest relative is Lagrangian remapping \citep{Lufkin1993}, where the fluid equations are solved in a Lagrangian frame to reduce advection errors, but the solution is interpolated (remapped) back onto an Eulerian grid. The underlying grid in our method is not strictly Eulerian any more, requiring additional fluxes due to cell-wall motion. Therefore, the effect of the grid motion can be integrated at the same time order as the fluid equations. The grid shape cannot change, which renders the method less flexible than moving-mesh codes \citep{Hopkins2015,2010Springel}. While dissipative properties can vary with time due to expansion or contraction of the underlying grid, they stay constant across the grid, in difference to adaptive mesh refinement techniques \citep{2000Fryxell,2002Teyssier,2004OShea,2009Cunningham,Stone2020}. \revise{The closest implementations to ours are JET and DISCO, two moving mesh codes which restrict to singular dimensions \citep{Duffell2013,Duffell2016}.}

\section{Summary}
We present an implementation of a co-scaling (expanding, contracting, or co-moving) grid for the magnetohydrodynamics code Athena++ \citep{Stone2020}. The method can be used to follow the evolution of a system over orders of magnitude in scale, as long as the underlying assumption of an existing flow symmetry persists. The scaling prescription ensures the preservation of cell aspect ratios.  The method's main strength lies in covering orders of magnitude in spatial scales for isotropically expanding or contracting systems, or for comoving systems, while keeping dissipative properties constant across the grid. It provides less flexibility than moving mesh codes, but it can be combined with Athena++'s native adaptive mesh refinement. 

 
\section{Acknowledgments}
We thank the University of North Carolina at Chapel Hill's Information Technology Services for providing the computational resources. 

We also thank Dr. Ellen Zweibel for her help in editing this paper and for encouraging the completion of this project. 
This work was aided by Dr. Zweibel's NSF grant AST-2007323. 
\revise{We thank the anonymous referee for a very constructive report, and especially for pointing out an inconsistency in an earlier version of our grid implementation.}

\bibliography{sample631}{}

\appendix
\section{Update Equation Derivation}
We show the derivation for both time independent and time dependent grids (including our co-scaling example) to highlight their differences, using a Cartesian grid as an example.

\subsection{Fixed Eulerian Grid Method }
\revise{For a static, time independent grid, the integration of the left hand side (LHS) of Eqn.\,\ref{eqn:PDE} over space gives}
\begin{equation}
   \int_{V} dV \left[ \frac{\partial \textbf{U}(\vec{x},t)}{\partial t} \right]
   =\frac{d }{d t} \int_{V} dV
    \left[ \textbf{U}(\hat{x},t) \right].
   \label{eqn:StaticLHS}
\end{equation}
\revise{Because the volume is independent of time, we can move the time derivative outside of the volume integral (See Eqn.\,\ref{eqn:RTTStatic}).}

\revise{
We then apply the same integration to the right hand side (RHS) of Eqn.\,\ref{eqn:PDE}. Considering a Cartesian coordinate system with unit vectors $\hat{x},\, \hat{y},\, \hat{z}$, the matrix of fluxes can be split into the directional quantities $(\textbf{F},\textbf{G},\textbf{H})$. This means the RHS is
\begin{equation}
    - \int_{V} dV \, \left[ \vec{\nabla}^T \cdot\left( \textbf{F}^T \hat{x} + \textbf{G}^T \hat{y} + \textbf{H}^T \hat{z}\right) \right].
    \label{eqn:StaticRHSSplitIntegral}
\end{equation}}

\revise{
By splitting the above volume integral into its constituent directions, the gradients in each direction can be removed. The integral of the flux divergence in the $\hat{x}$ direction is
\begin{equation}
     \int_{V} dV \left[ \frac{\partial  \textbf{F} }{\partial x} \right] \\=
     \iint dy dz \, \left[ \textbf{F}(x_{b},t) - \textbf{F}(x_{a},t)\right],
     \label{eqn:CartesianFluxExample}
\end{equation}
where $x_b$ and $x_a$ are the upper and lower bounds, respectively, of the volume $V$ in the $\hat{x}$ direction.}
\revise{
Using notation from \cite{Stone2008} and \cite{Stone2020}, the LHS and RHS can be written in a computationally usable form. The first step in using this notation is to consider solving the equation on a computational grid or mesh, where each direction $(x,y,z)$ is indexed by an integer $(i,j,k)$ and each cell in the grid has a unique tuple of these integers. Each cell has a volume which we define as $V_{i,j,k}$. With these assumptions, the volume average of the conservative values $\textbf{U}$ at time $t$ is
\begin{equation}
    \textbf{U}_{i,j,k} =  \frac{1}{V_{i,j,k}}\int_{V_{i,j,k}} \left[  \textbf{U}(x^1_i,x^2_j,x^3_k,t) \right] dV.
    \label{eqn:StaticAveU}
\end{equation}
Written with this notation, the LHS (Eqn.\,\ref{eqn:StaticLHS}) becomes
\begin{equation}
   V_{i,j,k}  \frac{d \textbf{U}_{i,j,k} }{d t}.
   \label{eqn:StaticLHSNumerical}
\end{equation} }
We also need to define the integral over the divergence of the fluxes. To discretize Eqn.\,\ref{eqn:CartesianFluxExample}, we can use the term
\revise{
\begin{multline}
     \textbf{F}_{i+\frac{1}{2},j,k}= \frac{1}{\Delta y_j \Delta z_k}\\
     \iint dy dz \, \left[ \textbf{F}(x_{i+\frac{1}{2}},y,z,t) \right],
    \label{eqn:StaticAveF}
\end{multline}
}
where $x_{i+\frac{1}{2}}$ is the upper bound of the cell centered (with respect to the $\hat{x}$ coordinate) on $x_i$. The $\Delta y_j$ and $\Delta z_k$ are the width of the cell in the $\hat{y}$ and $\hat{z}$ directions respectively. Eqn.\,\ref{eqn:StaticAveF} is the value of the flux of each conservative variable at the given wall of the cell. This notation can be used not only for $\textbf{F}$, but also for $\textbf{G}$ and $\textbf{H}$. Altogether, the RHS (Eqn.\,\ref{eqn:StaticRHSSplitIntegral}) can be written as
\revise{
\begin{multline}
    -\bigg[ \Delta y_j \Delta z_k \left(  
    \textbf{F}_{i+\frac{1}{2},j,k} - \textbf{F}_{i-\frac{1}{2},j,k} \right) \\
    + \Delta x_i \Delta z_k \left(
    \textbf{G}_{i,j+\frac{1}{2},k} - \textbf{G}_{i,j-\frac{1}{2},k}\right)\\
    + \Delta x_i \Delta y_j \left(
    \textbf{H}_{i,j,k+\frac{1}{2}} - \textbf{H}_{i,j,k-\frac{1}{2}}\right) \bigg].
    \label{eqn:StaticRHSNumerical}
\end{multline}
}
Combining the LHS and RHS, we get the update equation (Eqn.\,\ref{eqn:StaticUpdate}) for a discrete Cartesian grid. \revise{This is the final update equation and it is used in Athena++ to evolve the system \citep{Stone2008,Stone2020}.} The entire derivation assumes that the grid does not depend on time. The next section outlines how the update equation changes when cell positions and sizes depend on time.

\subsection{Time-Dependent Eulerian Grid Method}
The most important change to the static grid update equation (Eqn.\,\ref{eqn:StaticUpdate}) derivation in the co-scaling grid case comes from the Reynolds Transport Theorem, Eqn.\,\ref{eqn:RTT}. \revise{ Integrating Eqn.\,\ref{eqn:PDE} over space to make the LHS into Eqn.\,\ref{eqn:StaticLHS} is still valid in the moving grid method. However, we cannot simply move the time derivative outside of the volume integral. Instead, the LHS will be 
\begin{multline}
    \int_{V} dV \left[ \frac{\partial \textbf{U}(\vec{x},t)}{\partial t} \right] = \\
    \frac{d}{dt}\int_{V(t)} dV \left[  \textbf{U}(\vec{x},t) \right] 
    -\int_{B} dA (\vec{w}\cdot \hat{n})\textbf{U}.
     \label{eqn:DynamicLHSIntegral}
\end{multline}}
The above expression clearly indicates that there is an additional flux due to grid motion $\vec{w}$. This term can be directly incorporated to the RHS expression. Moving the extra term in Eqn.\,\ref{eqn:DynamicLHSIntegral} to the RHS, Eqn.\,\ref{eqn:StaticRHSSplitIntegral}, the RHS for a Cartesian grid becomes
\revise{
\begin{multline}
    -\Bigg[  \int_{V} dV \, 
    \left[  \vec{\nabla}^T \cdot\left( \textbf{F}^T \hat{x} + \textbf{G}^T \hat{y} + \textbf{H}^T \hat{z}\right) \right]\\
    - \int_{B} dA \, (\vec{w}\cdot \hat{n})\textbf{U} \Bigg].
    \label{eqn:DynamicRHSIntegral}
\end{multline}}
This is the most general formulation of the so-called `true flux' \citep{springel2011} for a time dependent grid. As seen in the fixed grid derivation, the volume integral of the fluxes becomes an \textit{area} average over each directional flux, $\textbf{F}$, $\textbf{G}$, and $\textbf{H}$. The RHS will be \revise{
\begin{multline}
    - \bigg[ \int_{V(t)} dV \left[ \frac{\partial  \textbf{F}}{\partial x} +\frac{\partial  \textbf{G} }{\partial y} +\frac{\partial  \textbf{H} }{\partial z} \right] \\
    - \sum^{6}_{m=1} \int_{A_m} dA \, w_m \textbf{U} \Bigg],
    \label{eqn:DynamicRHSCart}
\end{multline}}
where we have split the integration over the boundary $B$ into a sum of integrals over the 6 walls of the Cartesian volume element. The velocity \revise{$w_m = \vec{w}_m \cdot \hat{n}_m$} is evaluated on the wall in the direction of the normal vector to $A_m$.

Ignoring the $y$ and $z$ directions, the RHS is
\revise{
\begin{multline}
    - \,\bigg[ \int_{y(t)}\int_{z(t)} dy dz \bigg[ \textbf{F} (x_{i+\frac{1}{2}}) -  \textbf{F}(x_{i-\frac{1}{2}}) \\
    -\left[ w_x(x_{i+\frac{1}{2}}) \textbf{U}(x_{i+\frac{1}{2}}) -  w_x(x_{i-\frac{1}{2}}) \textbf{U}(x_{i-\frac{1}{2}}) \right] \bigg]\Bigg].
    \label{eqn:DynamicRHSXFlux}
\end{multline}}
The negative sign for the flux at the $x_{i-\frac{1}{2}}$ wall results from the normal vector $\hat{n} = -\hat{x}$ at that wall. Taking note of this relationship between $\hat{n}$ and $\hat{x}$ allows us to write the flux using the $x$ component of $\vec{w}$ at that wall, which is defined as $w_x(x_{i-\frac{1}{2}})$.  

\revise{
To formulate a new update equation, we need to define a wall flux for the various directions. We use $\textbf{W}$, $\textbf{V}$, and $\textbf{S}$ to denote the wall flux in the $x$, $y$, and $z$ directions respectively. As a result, the numerical term for the moving wall flux is
\begin{multline}
    \textbf{W}_{i+\frac{1}{2},j,k} =
    \frac{1}{\Delta y_j \Delta z_k}  \\
    \iint dy dz \,
    \left[ w_x(x_{i+\frac{1}{2}}(t)) \textbf{U}(x_{i+\frac{1}{2}}(t),t) \right].
    \label{eqn:DynamicsAveWallFlux}
\end{multline}
}

Using the notation above, the RHS is
\revise{
\begin{multline}
    -\bigg[ \Delta y_j \Delta z_k \Bigg(  \textbf{F}_{i+\frac{1}{2},j,k} -   \textbf{F}_{i-\frac{1}{2},j,k} - \textbf{W}_{i+\frac{1}{2},j,k}+
    \textbf{W}_{i-\frac{1}{2},j,k} \Bigg)  \\
    + \Delta x_i \Delta z_k \Bigg(
    \textbf{G}_{i,j+\frac{1}{2},k} -  \textbf{G}_{i,j-\frac{1}{2},k}  -\textbf{V}_{i,j+\frac{1}{2},k}+  \textbf{V}_{i,j-\frac{1}{2},k} \Bigg)\\
    + \Delta x_i \Delta y_j \Bigg(
    \textbf{H}_{i,j,k+\frac{1}{2}} -  \textbf{H}_{i,j,k-\frac{1}{2}}  -\textbf{S}_{i,j,k+\frac{1}{2}} +  \textbf{S}_{i,j,k-\frac{1}{2}}\Bigg)\bigg].
    \label{eqn:DynamicRHSNumerical}
\end{multline}
}

Combining the LHS and RHS that we have derived above, we get an update equation (Eqn.\,\ref{eqn:DynamicUpdate}) for a simulation with a time dependent grid. \revise{ Considering the time integration involved in solving the ordinary differential equation Eqn.~\ref{eqn:DynamicUpdate}, we also find a necessary volume change correction (see Eqn.~\ref{eqn:TimeInt2}).}

\end{document}